\documentclass[5p,final]{elsarticle}

\newcommand{\mathdefault}[1][]{}

\usepackage{amssymb}
\usepackage{amsmath}

\usepackage{mathdots} 
\usepackage{dsfont}
\usepackage{graphicx}
\usepackage{nicefrac}
\usepackage{booktabs}
\usepackage{tikz}
\usetikzlibrary{calc,math,patterns,shapes,backgrounds,patterns.meta, arrows.meta, positioning, fit}
\usepackage{pgf}
\usepackage{pgfplots}
\usepackage{todonotes} 
\setuptodonotes{inline}
\usepackage{tabularx}
\usepackage[boxed,noline,noend,ruled,linesnumbered]{algorithm2e}
\usepackage{xintexpr}
\usepackage{boxedminipage}
\usepackage{framed}
\usepackage{tcolorbox}
\usepackage{etoolbox}
\usepackage{xifthen}
\usepackage{mathtools}

\makeatletter
\def\ps@pprintTitle{%
 \let\@oddhead\@empty
 \let\@evenhead\@empty
 \let\@evenfoot\@oddfoot}
\makeatother
\usepackage{lipsum}
\usepackage{lineno}

\usepackage{xurl}
\usepackage{hyperref}
\usepackage[hyphenbreaks]{breakurl}

\usepackage{tikz,pgfplots,pgfplotstable,xcolor}
\usepackage{ifthen}
\usepackage{graphics}
\usepackage{subcaption}

\usepackage[numbers]{natbib}
\modulolinenumbers[5]
\hypersetup{colorlinks, breaklinks=true}
\usepackage[noabbrev,nameinlink,capitalize]{cleveref}
\pgfplotsset{compat=1.16}

\pgfplotsset{
	discard if not/.style 2 args={
		x filter/.append code={
			\edef\tempa{\thisrow{#1}}
			\edef\tempb{#2}
			\ifx\tempa\tempb
			\else
				
			\fi
		}
	},
	discard if/.style 2 args={
		x filter/.append code={
			\edef\tempa{\thisrow{#1}}
			\edef\tempb{#2}
			\ifx\tempa\tempb
				
			\else
			\fi
		}
	}
}
\pgfmathdeclarefunction{lg2}{1}{%
    \pgfmathparse{ln(#1)/ln(2)}%
}
\pgfmathdeclarefunction{lg10}{1}{%
    \pgfmathparse{ln(#1)/ln(10)}%
}

\newlength{\RoundedBoxWidth}
\newsavebox{\GrayRoundedBox}
\newenvironment{GrayBox}[1]%
   {\setlength{\RoundedBoxWidth}{.4\textwidth}
    \def\boxheading{#1}
    \begin{lrbox}{\GrayRoundedBox}
       \begin{minipage}{\RoundedBoxWidth}}%
   {   \end{minipage}
    \end{lrbox}
    \begin{center}
    \begin{tikzpicture}%
       \node(Text)[draw=black!20,fill=white,rounded corners,%
             inner sep=2ex,text width=\RoundedBoxWidth]%
             {\usebox{\GrayRoundedBox}};
        \coordinate(x) at (current bounding box.north west);
        \node [draw=white,rectangle,inner sep=3pt,anchor=north west,fill=white] 
        at ($(x)+(6pt,.75em)$) {\boxheading};
    \end{tikzpicture}
    \end{center}}

\newenvironment{defproblemx}[2][]{\noindent\ignorespaces%
                                \FrameSep=6pt%
                                \parindent=0pt%
                \vspace*{-1.5em}
                \ifthenelse{\isempty{#1}}{%
                  \begin{GrayBox}{\textsc{#2}}%
                }{%
                  \begin{GrayBox}{\textsc{#2}  parameterized by~{#1}}%
                }
                \begin{tabular*}{\textwidth}{@{\hspace{.1em}} >{\itshape} p{0.72cm} p{0.84\textwidth} @{}}%
            }{
                \end{tabular*}%
                \end{GrayBox}%
                \ignorespacesafterend
            }  

\newcommand{\defproblem}[3]{
  \begin{defproblemx}{#1}
    Input:  & #2 \\
    Task: & #3
  \end{defproblemx}
}%

\begin{document}
\begin{frontmatter}
	\title{Arcee: An \textsc{OCM}-Solver}
	\author[1]{Kimon Boehmer}
    \author[2]{Lukas Lee George}
    \author[3]{Fanny Hauser}
    \author[4]{Jesse Palarus\corref{cor1}}
    \cortext[cor1]{Corresponding author}
    \ead{j.palarus@gmail.com}
    \address[1]{Université Paris-Saclay, kimon.boehmer@ens-paris-saclay.fr}
    \address[2]{TU Berlin, l.l.george@gmx.de}
    \address[3]{TU Berlin, f.hauser@tu-berlin.de}
    \address[4]{TU Berlin, j.palarus@tu-berlin.de}
 \begin{abstract}
The 2024 PACE Challenge focused on the \textsc{One-Sided Crossing Minimization} (OCM) problem, which aims to minimize edge crossings in a bipartite graph with a fixed order in one partition and a free order in the other. 
We describe our OCM solver submission that utilizes various reduction rules for OCM and, for the heuristic track, employs local search approaches as well as techniques to escape local minima. The exact and parameterized solver uses an ILP formulation and branch \& bound to solve an equivalent \textsc{Feedback Arc Set} instance.
 \end{abstract}
\begin{keyword}
    PACE 2024, One-Sided Crossing Minimization, OCM
\end{keyword}
\end{frontmatter}


\section{Introduction} 
In the Parameterized Algorithms and Computational Experiments (PACE) Challenge of 2024, the problem of interest was \textsc{One-Sided Crossing Minimization} (OCM).
In this problem, we are given a bipartite graph with vertex partitions $A$ and $B$ which are drawn horizontally and in parallel.
Additionally, we are given a fixed linear order for the vertices in $A$.
The goal is to find an ordering of the vertices in $B$ that minimizes the total number of edge crossings when all edges are drawn with straight lines.

Algorithms for OCM are used for drawing hierarchical graphs \cite{battista1998graph} or producing row-based VLSI layouts \cite{sechen2012vlsi,stallmann2001heuristics}.

\defproblem{One-Sided Crossing Minimization}
{A bipartite graph $G =((A \cup B),E)$, and a linear order of $A$.}
{Find an linear ordering of the vertices in $B$ that minimizes the total number of edge crossings in a straight-line drawing of $G$ with $A$ and $B$ on two parallel lines, following their linear order.}

\subsection{Related Work}
\citet{eades1994edge} showed that OCM is \textbf{NP}-hard and \citet{dobler2023note} further strengthened this result by showing that it remains \textbf{NP}-hard on trees. 
Positively, several fixed-parameter tractable algorithms have been proposed \cite{dujmovic2006fixed,dujmovic2008parameterized,FPT_OCM}. \citet{FPT_OCM} showed that OCM has a polynomial kernel.
When lifting the exactness constraints, two well known and simple heuristics are the \emph{median heuristic} which was introduced by \citet{eades1994edge} and the \emph{barycenter heuristic} which was introduced by \citet{PenaltyGraph}.
The median heuristic places each vertex in $B$ at the median of the positions of its neighboring vertices in $A$. Its running time running time is $\mathcal{O}(n \log n)$.
The barycenter heuristic has a running time of $\mathcal{O}(m+n \log n)$ and is similar to the median heuristic with the difference that each vertex of $B$ is placed on the average of the positions of its neighboring vertices in $A$

\citet{eades1994edge} also showed that the median heuristic is a 3-approximation. Later \citet{nagamochi2005improved} proposed a 1.4664-approximation algorithm.

\subsection{PACE}
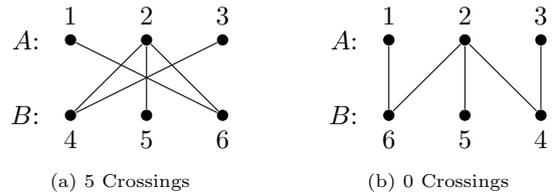
\begin{figure}[t]
    \centering
    \begin{subfigure}[t]{0.2\textwidth}
        \centering
        \begin{tikzpicture}[node distance=1cm and 0.5cm]
            \node (v1) at (0,1) [circle, fill=black, inner sep=1.5pt, label=above:1] {};
            \node (v2) at (1,1) [circle, fill=black, inner sep=1.5pt, label=above:2] {};
            \node (v3) at (2,1) [circle, fill=black, inner sep=1.5pt, label=above:3] {};

            \node (v4) at (0,0) [circle, fill=black, inner sep=1.5pt, label=below:4] {};
            \node (v5) at (1,0) [circle, fill=black, inner sep=1.5pt, label=below:5] {};
            \node (v6) at (2,0) [circle, fill=black, inner sep=1.5pt, label=below:6] {};

            \draw (v2) -- (v6);
            \draw (v1) -- (v6);
            \draw (v2) -- (v5);
            \draw (v2) -- (v4);
            \draw (v3) -- (v4);
            
            \node[left=0.2cm of v1] {$A$:};
            \node[left=0.2cm of v4] {$B$:};
            
        \end{tikzpicture}
        \caption{5 Crossings}
    \end{subfigure}
    \hspace{8pt}
    \begin{subfigure}[t]{0.2\textwidth}
        \centering
        \begin{tikzpicture}[node distance=1cm and 0.5cm]
            \node (v1) at (0,1) [circle, fill=black, inner sep=1.5pt, label=above:1] {};
            \node (v2) at (1,1) [circle, fill=black, inner sep=1.5pt, label=above:2] {};
            \node (v3) at (2,1) [circle, fill=black, inner sep=1.5pt, label=above:3] {};

            \node (v4) at (2,0) [circle, fill=black, inner sep=1.5pt, label=below:4] {};
            \node (v5) at (1,0) [circle, fill=black, inner sep=1.5pt, label=below:5] {};
            \node (v6) at (0,0) [circle, fill=black, inner sep=1.5pt, label=below:6] {};

            \draw (v2) -- (v6);
            \draw (v1) -- (v6);
            \draw (v2) -- (v5);
            \draw (v2) -- (v4);
            \draw (v3) -- (v4);

            \node[left=0.2cm of v1] {$A$:};
            \node[left=0.2cm of v6] {$B$:};
        \end{tikzpicture}
        \caption{0 Crossings}
    \end{subfigure}
    \caption{An OCM instance with different orderings of the free vertices and the corresponding number of edge crossings.}
    \label{fig:combined_figure}
\end{figure}
The Parameterized Algorithms and Computational Experiments Challenge (PACE) \footnote{\url{https://pacechallenge.org/2024/}} was first held in 2016 with the goal to deepen the relationship between parameterized algorithms and practice.
Each year an \textbf{NP}-hard problem is given and the goal is to program a solver for this problem. Previous challenges included problems like \textsc{Tree\allowbreak width}, \textsc{Feedback Vertex Set}, \textsc{Cluster Editing} or \textsc{Vertex Cover}. This years challenge was announced in November 2023 and the submission deadline was in June 2024. The challenge consisted of three tracks. In the Exact Track, the solver has 30 minutes to find the optimal solution for the problem instance. In the Heuristic Track, the goal is to find the best possible solution within 5 minutes. The last track is the Parameterized Track, which is similar to the exact track because the solver has 30 minutes to find the optimal solution but additionally all instances have \emph{small cutwidth}.
It was required that all solvers are single threaded and each solver is limited to 8GB of memory.
The solvers were tested on 200 instances for each track (100 of those were known during the competition) and we used those to test our solvers while developing them.
In \cref{fig:combined_figure} we can see an example instance with the PACE naming scheme and a corresponding optimal solution to this instance.

\section{Preliminaries} \label{sec:preliminaries}
We make use of the usual definitions for graphs $G = (V, E)$, bipartite graphs $ G = ((A \cup B), E)$ and directed graphs $G = (V, E)$.
We use $N(v)$ for a vertex $v \in V$ to denote its open neighborhood and $N(U)$ for a set $U$ to denote the neighborhood union of all $u \in U$ as $N(U) = \bigcup_{u\in U} N(u)$.
We call the set $B$ of an OCM instance the free vertices set or free set and $A$ the fixed vertices set or fixed set.

Usually, exact and heuristic solvers for OCM will first require the computation of the so-called \emph{crossing matrix} $M$ or \emph{crossing numbers} \cite{FPT_OCM, junger20022, interval_graph_FPT_OCM, OCM_sifting}.
An entry $c_{uv} \in M$ with $u, v \in B$ denotes the number of edge crossings between edges incident to $u$ and edges incident to $v$, when $u$ appears before $v$ in the ordering.
In other words, this is the number of pairs of fixed vertices $u', v' \in A$ for which $u'$ precedes $v'$ in the fixed order and $u' \in N(v)$ and $ v' \in N(u)$.

\begin{figure}
    \centering
        \begin{tikzpicture}[node distance=1cm and 0.5cm]
            \node (v1) at (0,1) [circle, fill=black, inner sep=1.5pt, label=above:1] {};
            \node (v2) at (0.5,1) [circle, fill=black, inner sep=1.5pt, label=above:2] {};
            \node (v3) at (1.25,1) [circle, fill=black, inner sep=1.5pt, label=above:3] {};
            \node (v4) at (1.5,1) [circle, fill=black, inner sep=1.5pt, label=above:4] {};
            \node (v5) at (2.5,1) [circle, fill=black, inner sep=1.5pt, label=above:5] {};
            \node (v6) at (4,1) [circle, fill=black, inner sep=1.5pt, label=above:6] {};
            \node (v7) at (4.25,1) [circle, fill=black, inner sep=1.5pt, label=above:7] {};
            \node (v8) at (4.5,1) [circle, fill=black, inner sep=1.5pt, label=above:8] {};
            \node (v9) at (5.5,1) [circle, fill=black, inner sep=1.5pt, label=above:9] {};

            \node (v10) at (1.33,0) [circle, fill=black, inner sep=1.5pt, label=below:10] {};
            \node (v11) at (2.5,0) [circle, fill=black, inner sep=1.5pt, label=below:11] {};
            \node (v12) at (4.25,0) [circle, fill=black, inner sep=1.5pt, label=below:12] {};

            \draw (v1) -- (v12);
            \draw (v2) -- (v12);
            \draw (v3) -- (v10);
            \draw (v4) -- (v10);
            \draw (v5) -- (v11);
            \draw (v6) -- (v12);
            \draw (v7) -- (v12);
            \draw (v8) -- (v12);
            \draw (v9) -- (v10);

            \node[left=0.2cm of v1] {$A$:};
            \node[left=0.2cm of v10] {$B$:};
        \end{tikzpicture}
        \caption{OCM instance whose penalty graph contains a cycle.}
        \label{fig:ocm_cycle}
        \end{figure}
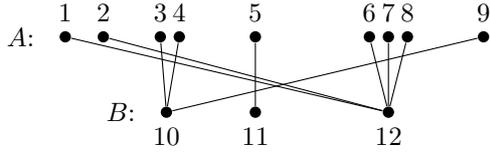

            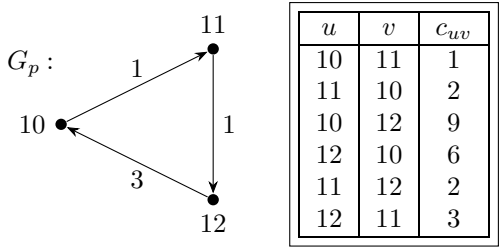
\begin{figure}
            \centering
\begin{tikzpicture}[
    ->, >=Stealth, node distance=2cm, on grid,
]
    \node[draw, anchor=west] (table) at (3, 0) {
        \begin{tabular}{|c|c|c|}
            \hline
            $u$ & $v$ & $c_{uv}$ \\
            \hline
            10 & 11 & 1 \\            
            11 & 10 & 2 \\
            10 & 12 & 9 \\
            12 & 10 & 6 \\
            11 & 12 & 2 \\
            12 & 11 & 3 \\
            \hline
        \end{tabular}
    };

    \node (v10) at (0,0) [circle, fill=black, inner sep=1.5pt, label=left:10] {};
    \node (v11) at (2,1) [circle, fill=black, inner sep=1.5pt, label=above:11] {};
    \node (v12) at (2,-1) [circle, fill=black, inner sep=1.5pt, label=below:12] {};

    \path[->]
        (v10) edge node[midway, above] {1} (v11);
    \path[->]
        (v11) edge node[midway, right] {1} (v12);
    \path[->]
        (v12) edge node[midway, below] {3} (v10);

    \node (GP) at (-0.4, 0.8) {$G_p$\::};
\end{tikzpicture}
\caption{Penalty graph $G_p$ and crossing numbers of the instance in \cref{fig:ocm_cycle}.}
\label{fig:penalty_graph}
\end{figure}

The \emph{penalty graph} $G_p$ of an OCM instance with free vertices set $B$ is a weighted directed graph $(B,E)$ where $E:=\{(u,v) \in B^2 \mid c_{uv}<c_{vu}\}$ and edge weights $w((u,v))=c_{vu}-c_{uv}$. \citet{PenaltyGraph} observed a connection between OCM and the Weighted Feedback Arc Set, which we will refer to as Feedback Arc Set or FAS in the following, of the penalty graph: An optimal ordering of the vertices $B$ for OCM is equal to a topological ordering when an optimal Feedback Arc Set is removed from the penalty graph.

\defproblem{Weighted Feedback Arc Set}
{A directed graph $G =(V, E)$ and a function $w: E \mapsto \mathds{N}$}
{Find a set of edges $X \subseteq E$ such that $G_X = (V, E \setminus X)$ is acyclic while minimizing $\sum_{x \in X}w(x)$.}

Intuitively the penalty graph is generated by orienting every pair such that the crossing number is minimized and then resolves cycles in that order with FAS paying exactly the delta of a pair's crossing numbers in order to remove that edge in the penalty graph.
The instance in \cref{fig:ocm_cycle} can be optimally solved using the penalty graph in \cref{fig:penalty_graph} by removing e.g. edge $(10, 11)$. Thus, the optimal ordering provided via the topological sort of the penalty graph with edge $(10, 11)$ removed would be $11$ before $12$ before $10$.

In the following, we consider OCM instances and graphs to be \textit{large} if the solver opts not to generate and store their crossing matrix and penalty graph due to memory limitations.
In our submitted solver all instances with more than $10,000$ free vertices are considered large.

\section{Data Reduction} \label{sec:data_reduction}
Before solving any instance with the heuristic or exact techniques we try to split it up into a set of subinstances and reduce them.
The following methods, consisting of graph splitting and reduction rules, are employed in the heuristic, exact and parameterized track.

\subsection{Graph Splitting}
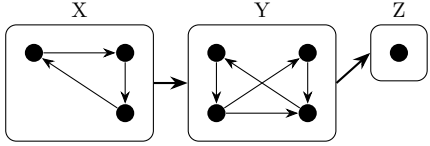
\begin{figure}[t]
        \centering
        \begin{tikzpicture}[->, >=Stealth, node distance=1cm, on grid, auto, scale=0.8, transform shape]
            \tikzset{
                component/.style = {draw, rectangle, rounded corners, inner sep=10pt, fit=#1},
                node/.style = {circle, fill=black, inner sep=3pt},
            }

            \node[node] (A1) {};
            \node[node, right=1.5cm of A1] (A2) {};
            \node[node, below=1cm of A2] (A3) {};

            \path[->]
                (A1) edge (A2)
                (A2) edge (A3)
                (A3) edge (A1);

            \node[component=(A1)(A2)(A3), label=above:X, inner sep=7pt] (boxA) {};

            \node[node, right=1.5cm of A2] (B4) {};
            \node[node, below=1cm of B4] (B5) {};
            \node[node, right=1.5cm of B4] (B6) {};
            \node[node, below=1cm of B6] (B7) {};

            \path[->]
                (B4) edge (B5)
                (B5) edge (B6)
                (B6) edge (B7)
                (B7) edge (B4)
                (B5) edge (B7);

            \node[component=(B4)(B5)(B6), label=above:Y, inner sep=7pt] (boxB) {};

            \node[node, right=1.5cm of B6] (C6) {};

            \node[component=(C6), label=above:Z, inner sep=7pt] (boxC) {};

            \path[->, thick]
                (boxA.east) edge (boxB.west)
                (boxB.east) edge (boxC.west);
        \end{tikzpicture}
        \caption{Strongly connected components of a penalty graph.}
        \label{fig:strongly_connected_components}
    \end{figure}

Our go-to approach in order to split an instance's graph relies on the penalty graph $G_p$.
We observe that we can solve each strongly connected component of the instance's penalty graph individually.
Concatenating their solutions in the topological order of $G_p$'s strongly connected components (visualized in \cref{fig:strongly_connected_components}) yields a correct linear order for $B$.
The order is optimal if all of the penalty graph's strongly connected components were solved optimally, due to the topological sort ordering components such that their crossings are minimized.

\begin{figure}[h]
        \centering
        \begin{tikzpicture}[node distance=1cm and 0.5cm]
            \node (v1) at (0,1) [circle, fill=black, inner sep=1.5pt, label=above:1] {};
            \node (v2) at (2,1) [circle, fill=black, inner sep=1.5pt, label=above:2] {};
            \node (v3) at (3,1) [circle, fill=black, inner sep=1.5pt, label=above:3] {};
            \node (v4) at (5,1) [circle, fill=black, inner sep=1.5pt, label=above:4] {};

            \node (v5) at (0,0) [circle, fill=black, inner sep=1.5pt, label=below:5] {};
            \node (v7) at (1,0) [circle, fill=black, inner sep=1.5pt, label=below:7] {};
            \node (v9) at (2,0) [circle, fill=black, inner sep=1.5pt, label=below:9] {};
            \node (v6) at (3,0) [circle, fill=black, inner sep=1.5pt, label=below:6] {};
            \node (v8) at (4,0) [circle, fill=black, inner sep=1.5pt, label=below:8] {};
            \node (v10) at (5,0) [circle, fill=black, inner sep=1.5pt, label=below:10] {};

            \draw (v1) -- (v5);
            \draw (v1) -- (v7);
            \draw (v1) -- (v9);
            \draw (v2) -- (v5);
            \draw (v2) -- (v7);
            \draw (v2) -- (v9);

            \draw (v3) -- (v6);
            \draw (v3) -- (v8);
            \draw (v3) -- (v10);
            \draw (v4) -- (v6);
            \draw (v4) -- (v8);
            \draw (v4) -- (v10);
            
            \node[left=0.2cm of v1] {$A$:};
            \node[left=0.2cm of v5] {$B$:};
        \end{tikzpicture}
        \caption{OCM instance splittable by partitioning the set of free vertices into $\{\{5, 7, 9\}, \{6, 8, 10\}\}$.}
        \label{fig:OCM_instance}
\end{figure}
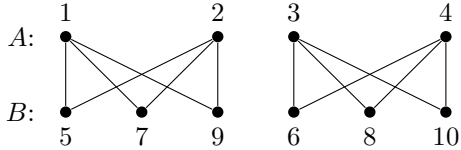

However, the penalty graph approach is only feasible for small graphs.
Therefore, we rely on a simpler method for large graphs where we try to split the graph by partitioning the free vertices $B$ into non-empty subsets $\mathcal{B} = \{B_1, B_2, \dots, B_k\}$ such that there are no vertices $u, v, w \in A$ with $u$ before $v$ before $w$ in the fixed order of $A$ and sets $B_i, B_j \in \mathcal{B}$ with $u, w \in N(B_i)$ and $v \in N(B_j)$.
In other words: The neighborhood intervals of the elements of $\mathcal{B}$ do not overlap in the set of fixed vertices.
An example OCM instance that can be split via partitioning is \cref{fig:OCM_instance} where one can observe that the partition of vertices in $B$ $\{5,7,9\}$ has to be ordered before $\{6,8,10\}$.
We can split a graph into the induced subgraphs of each partition element and potentially further split these subgraphs with the aforementioned penalty graph splitting approach.
Again, optimal solution orders for each subgraph can be concatenated to an optimal overall order for the entire instance.
This time they are ordered by their neighborhood intervals: Every vertex in $B_i$ is ordered before every vertex in $B_j$ if for all vertices $u \in N(B_i)$ and all vertices $v \in N(B_j)$ $u$ comes before $v$ in the given linear order of $A$ for all $B_i, B_j \in \mathcal{B}$.
In both methods isolated vertices in $B$ can be inserted into the overall solution order arbitrarily, because they do not have an effect on the number of crossings.

\subsection{Reduction Rules}
We mostly apply data reductions proposed by \citet{FPT_OCM}. In particular their rules \texttt{RR1}, \texttt{RR2}, \texttt{RRLO1} in unmodified form and a modified version of their \texttt{RRlarge} rule that accounts not only for an upper bound but also for the trivial lower bound described by \citet{OCM_lower_bound}.

Reduction rules \texttt{RR1}, \texttt{RR2} and \texttt{RRlarge} work by creating a partial ordering $p$ of the vertices in $B$ that is used by rule \texttt{RRLO1} to entirely remove vertices from the graph or fix the already ordered pairs in the solution.
For every $a, b \in B$ with $a \neq b$:
\begin{itemize}
\item Rule \texttt{RR1} commits $a$ before $b$ in $p$ if $c_{ab} = 0$. Dujmovic et al. additionally require $c_{ba} > 0$, which we do not have to enforce, because there is no parameter to account for in our case.
\item Rule \texttt{RR2} commits $a$ before $b$ arbitrarily if $N(a) = N(b)$.
\item The originally proposed \texttt{RRlarge} rule commits $a$ before $b$ if $c_{ba} > \textit{upper\_bound}$. The upper bound can be provided by a fast heuristic. Putting $b$ before $a$ in $p$ will result in a worse solution than the upper bound's solution.
\item Rule \texttt{RRLO1} eventually removes vertices from a graph if their final position in the solution output is fully determined by $p$. We can store the position and later add the vertex to the solution.
\end{itemize}
The trivial lower bound for any OCM instance can be found by summing up $min(c_{ab}, c_{ba})$ over all pairs $a,b \in B$.
We make use of that lower bound by incorporating it in the \texttt{RRlarge} rule whose modified version commits $a$ before $b$ if $c_{ba} + \textit{lower\_bound} - c_{ab} > \textit{upper\_bound}$.
The modified version is still correct.
Assume that an instance fulfills the inequality of our modified \texttt{RRlarge} rule.
Then putting $b$ before $a$ results in a solution with a number of total crossings $c_{ba}$ plus the lower bound of the instance without the pair $(a, b)$.
It must hold that $c_{ab} < c_{ba}$ as $c_{ab} > c_{ba}$ contradicts $\textit{lower\_bound} \leq \textit{upper\_bound}$.
Therefore, we can correct the lower bound to one without the pair $(a, b)$ by subtracting $c_{ab}$ so the solution with $b$ before $a$ results in $c_{ba} + \textit{lower\_bound} - c_{ab}$ total crossings which is worse than the upper bound solution.

Bringing everything together, we first run our graph splitting methods depending on the size of the instance and go on to use data reduction rules on each subinstance.
We apply \texttt{RR1}, \texttt{RR2} and our modified \texttt{RRlarge} rule exhaustively then add all transitive pairs to $p$ and, finally, remove vertices using \texttt{RRLO1}. Additionally we save every partially ordered pair $a,b \in B$ to make use of it in the exact and heuristic solvers.  

\section{Heuristic Solver} \label{sec:heuristic}
We use different approaches to find a heuristic solution depending on the size of the graph. For small graphs the repeated application of our methods leads to better results. However, on large graphs even a single application may stress the resource limit.
The first step in our heuristic for small graphs is to compute an initial order with the median heuristic. We will interpret the orderings of $A$ and $B$ as positions ranging from $1$ to $|A|$ and from $1$ to $|B|$, respectively. If two vertices are assigned the same position, we use an arbitrary tie breaker.

\subsection{Local Search}
\subparagraph{Sifting}
To improve an order, we use \emph{sifting}, which was first introduced by \citet{rudell1993dynamic}.
One vertex is taken at a time and placed at the position in the order which minimizes the total number of crossings. We do this exhaustively for all free vertices of the graph, but choose the sequence in which we sift the vertices randomly.
After applying sifting on the order computed by the median heuristic, we apply it on random orders and store the order with the fewest crossings.

\subparagraph{Swapping}
To escape local optima, we use \emph{force swapping} if no better solution was found for 592 iterations\footnote{The exact values were found with SMAC3 using a subset of the public instances for training \cite{SMAC}} of using a random order with sifting.
Force swapping is done by choosing two vertices from the free vertex set and swapping their positions in the order. Then, we use sifting to improve the order while requiring that the relative position of the two vertices remains unchanged. 
We iterate through all vertices in a random order, swapping each vertex with its immediate right neighbor within the current best ordering. Subsequently, we increment the distance of the swapped vertices in each iteration by 9 until the distance is greater than 90.
\subsection{Large Graphs}
For large graphs, we first compute an order with the median heuristic and try to improve it by swapping neighboring vertices in the order until $10\%$ of the available time is used or no swap of two neighboring vertices improves the solution. We repeat the same with the barycenter heuristic. We then check the number of crossings for both solutions and only work with the better solution in the following. Then, we use a variation of sifting to improve the best order found so far. Instead of checking all possible positions for each vertex in the order, we stop the search for the best insertion position of a vertex if the total number of crossings increases by $20,000$ or the vertex would be moved more than $2,000$ positions from its current position. These measures are mainly incorporated so that every vertex has a chance to be looked at in the time limit at least once even on very large graphs.
We sift the vertices until the time limit is reached.

\section{Exact Solver} 
For our exact solvers, we mainly use the fact from \cref{sec:preliminaries} that the solution for OCM is equivalent to a topological sorting from the corresponding penalty graph when an FAS is removed from the edge set. This implies that we will mainly focus on a way to find for a given directed graph $G=(V,E)$ and weight function $w:E\to \mathbb{N}$ a minimal FAS.

\subsection{ILP Solvers}

Integer Linear Programming (ILP) formulations are widely employed in the literature to obtain exact solutions for NP-hard problems \cite{ILP1, ILP2, ILP3}.
For our minimal FAS problem, we consider two approaches.

The first approach, proposed by \citet{ILPRef1}, introduces a binary variable $m_{u,v}$ for each node pair $u,v \in V, u \neq v$.
These variables encode a linear ordering such that $m_{u,v} = 1$ if and only if $u$ precedes $v$ in the ordering.
The ILP is formulated as follows:
\begin{align*}
    \min &\sum_{(v, u)\in E} m_{u,v} w(v, u)  \\
     \text{s.t. }& m_{u,v}\in \{0,1\} & &\text{ for all }   u, v \in V, u\neq v \\
      &m_{u,v} = 1 - m_{v,u} &  &\text{ for all }   u, v \in V, u\neq v \\
      &m_{u,v} + m_{v, w} - m_{u, w} \leq 1 &  &\text{ for all }   u, v, w \in V \\
      &&&                                       \text{ with } u\neq v, v \neq w, u \neq w
\end{align*}
The second constraint ensures symmetry in the resulting linear ordering, while the third constraint enforces transitivity.
Since this ILP solver finds a linear ordering with its variables, we will call it the linear ordering or linear ILP in the following.

An alternative ILP formulation, introduced by \citet{ILPRef2}, exploits the fact that removing at least one edge from every cycle yields an FAS. This leads to the following formulation:
\begin{align*}
      \min &\sum_{e \in E} w(e) \cdot y_e  \\
    \text{s.t. }& y_e \in \{0,1\} & &\text{ for all }   e \in E \\
      &\sum_{e \in C} y_e \geq 1 &  &\text{ for all } C\in \mathcal{C}
\end{align*}
Here, for each $e \in E$, we have a binary variable $y_e$ where $y_e=1$ if and only if edge $e$ should be removed from the graph. In the following, we will refer to this approach as cycle ILP.
\subsection{Row Generation}
The linear ordering approach utilizes $\mathcal{O}(n^2)$ variables and $\mathcal{O}(n^3)$ constraints. The high number of constraints can lead to substantial memory usage and numerical instability. In our experiments, we encountered instances with fewer than $100$ nodes where ILP solvers produced incorrect solutions due to numerical issues.

In contrast, the cycle-based approach employs $\mathcal{O}(m)$ variables and $\mathcal{O}(c)$ constraints, where $c$ is the number of cycles in the graph. For a complete graph, this results in $\mathcal{O}(n!)$ constraints in the worst case. Besides the aforementioned issues for the linear ordering ILP, even enumerating all these cycles is computationally hard.

To address these challenges, we employ a row generation technique. This approach iteratively adds constraints until a valid solution is found, mitigating memory usage and numerical stability issues.

For the linear ordering formulation, we initially remove all transitive constraints and solve the ILP. We then verify if the variables form a valid linear ordering. If not, we add only those transitive constraints that contradict the current solution. This process is repeated until a valid solution is obtained.

For the cycle-based formulation, we need not only to generate constraints but also to lazily generate cycles, as enumerating all cycles is computationally too expensive even for small graphs. We adapt an approach proposed by \citet{ILPSolver}, resulting in the algorithm presented in \cref{alg:exact_solver}.
\begin{algorithm}[t]
	\SetAlgoLined
	\KwIn{A bipartite graph $G = (A,B)$ and a linear ordering of $A$}
	calculate an ordering of $B$ using our heuristic from \cref{sec:heuristic} for 1 minute\;
	calculate the penalty graph $G_P = (V, E_{original})$\;
	$\mathcal{C} \gets \emptyset$\;
	\For{every edge $\{u,v\}$ that contradicts our heuristic}{
		add a shortest cycle containing $\{u,v\}$ to $\mathcal{C}$ (if it exists)\;
	}
	$X \gets$ solution from the partial ILP using $\mathcal{C}$\;
	go to step (4) using the graph $G'=(V, E_{original}\setminus X)$\;
	Repeat this until:
	\begin{enumerate}
		\item the size of the solution $X$ matches the heuristic\\ costs or
		\item $G'$ can be sorted topologically.
	\end{enumerate}
	\caption{Exact solver using lazy cycle generation}
	\label{alg:exact_solver}
\end{algorithm}
Our main modification to the original algorithm is in step 4. The paper suggests to recalculate an FAS heuristic for the current graph $G'$ in each iteration. Instead we utilize our initial heuristic and search for all edges that contradict it. This approach serves as a heuristic for graph $G'$ and has superior performance in our tests compared to recalculating an FAS heuristic at each iteration. We think that this improvement comes from our heuristic's ability to find correct solutions in all test instances. Because of our approach, we then force our partial cycle set $\mathcal{C}$ to include only those cycles necessary for the heuristic solution to be verified.
\subsection{Branching}
\label{sec:branching}
In addition to our ILP-based approach, we implemented a branch and bound algorithm. The main structure remains consistent with \cref{alg:exact_solver}, but instead of employing an ILP solver in line 6, we utilize a branching technique to solve the FAS problem with the partial cycle matrix $\mathcal{C}$. For this we use the observation that at least one edge must be removed from every cycle, leading to the following branching rule:

\noindent Search for a cycle $C$ with no selected edge:

For each edge $e \in C$: Select $e$ and make a recursive call \\

\noindent While this algorithm is not fixed-parameter tractable with respect to the solution size due to the unbounded number of recursive calls in each step, it performs well in practice. This efficiency is attributed to our strategy in line 5 of \cref{alg:exact_solver}, where we consistently add only the shortest cycles, resulting in relatively short cycles for branching.

To enhance the efficiency of our branch and bound algorithm, we add a packing lower bound. This is computed by searching for a set of edge-disjoint cycles $\mathcal{P}$. A simple lower bound can then be calculated as:
\begin{align*}
    \sum_{C \in \mathcal{P}} \min_{e \in C} w(e)
\end{align*}

We further improve this lower bound by allowing cycles to share edges in special cases. The procedure is as follows: We initialize a lower bound counter with 0. Then, we iterate through every cycle $C$ and search for the edge $e_{\min}$ with the smallest weight in $C$. We increment our lower bound by $w(e_{\min})$ and subtract $w(e_{\min})$ from the weight of each $e \in C$.

To cut branches more effectively, we implemented a local search heuristic based on the work of \citet{SetCover}. The core idea of this approach is to greedily select edges based on a random probability until a solution is found. The algorithm then iteratively improves this solution by randomly removing some edges from the current solution and reconstructing a solution using the randomized greedy strategy. This process is repeated for a fixed number of iterations, after which the best solution found is used as an upper bound.
\section{Parameterized Solver}
In the parameterized track of the 2024 PACE Challenge instances included an additional cutwidth parameter.
\citet{cutwidth} defines the cutwidth of a graph $G = (V, E)$ using an injective function $\pi: V \mapsto \mathds{N}$ called a numbering which denotes the order of vertices $G$ drawn on a straight line.
Then the cutwidth $cw$ of $G$ is:
$$cw(G) = \min_{\pi}\max_{i\in \mathds{N}}|\{\{u,v\} \in E : \pi(u) \leq i < \pi(v)\}|$$
Thus, the cutwidth of a graph is the maximum number of edges from an earlier to a later partition when drawing all vertices on a straight line in the order of a minimizing numbering.
The example in \cref{fig:ocm_cycle} has cutwidth $3$ by putting vertices in the order $11, 3, 7, 12, 6, 2, 1, 5, 10, 9, 8, 4$\footnote{The cutwidth was calculated using \url{https://github.com/lucas-t-reis/minimum-cutwidth}}.
\citet{cutwidth_lower_bound_cr} show that cutwidth is a lower bound to the crossing number of a graph.

The core OCM problem definition does not change for the parameterized track.
However, the cutwidth of the instance graph and the numbering that witnesses it are provided as additional input.

Nevertheless, we do not employ any techniques making use of the parameter or numbering and submitted a version of our exact solver in this track with a minor adjustment.
Due to the ILP solver's startup overhead we are, instead, solving small instances with an upper bound less than $10$ using the branch and bound algorithm from \cref{sec:branching}.

\section{Implementation Details}
To achieve high performance while maintaining the convenience features of modern programming languages, we implemented all code in C++17. Additionally this section details some of the optimization techniques we employed to further enhance performance.
\subsection{Graph data structure}
The core graph data structure comprises two adjacency lists: One for neighbors of set $A$ and another for neighbors of set $B$. These lists are sorted by vertex index, enabling us to calculate the crossing number of two nodes $u,v \in V$ in $\mathcal{O}(\min(|N(u)|, |N(v)|))$ time, instead of the $\mathcal{O}(|N(u)||N(v)|)$ complexity of a naive approach.

Furthermore for graphs with fewer than 10,000 nodes, we initialize a crossing matrix $M$. This matrix stores the crossing number $c_{u,v}$ for every pair $u,v \in B$ in the entry $M_{u,v}$. Additionally, we calculate a matrix $M' = M - M^\top$, which directly stores the values $c_{u,v} - c_{v,u}$. This matrix accelerates the sifting algorithm by eliminating the calculation of the subtractions and especially improving cache alignment.
\subsection{Fast crossing calculation}
The calculation of crossings for a given ordering is crucial, particularly for the heuristic algorithm to compare two potential solutions with each other. While the calculation via the crossing matrix suffices for small graphs, larger graphs require a more sophisticated approach. For this we implemented an algorithm inspired by \citet{Segtrees}, which iterates through each vertex in set $A$, considering its neighbors in $B$ ordered according to the current solution. For each neighbor, it counts the crossings created with edges incident to previously processed neighbors, efficiently utilizing a segment tree to maintain cumulative counts and perform range sum queries. This method allows us to calculate the number of crossings in $\mathcal{O}(\sum_{a \in A} |N(a)| (\log(|N(a)|) + \log(|B|)))$ time.
\subsection{Fast transitive closure}
Computing the transitive closure of the current partial order is necessary before applying certain reduction rules, like e.g. \texttt{RRLO1}. However, a naive Floyd-Warshall algorithm can be time-consuming for larger instances. We optimize this process by leveraging the fact that a partial order always corresponds to a directed acyclic graph (DAG). This property allows us to calculate a topological ordering of the DAG.

Our approach involves traversing the topological order in reverse, propagating reachability information. To further enhance performance, we utilize C++'s \texttt{bitset} class for efficient logical operations on these reachability sets. This combination of topological ordering and bitset operations significantly reduces the computation time for the transitive closure, especially for larger graph instances.
\section{Experiments} 
In the following section, we will discuss the results of our experiments and look at the final results of the PACE challenge.
For this we ran all experiments on an \texttt{Intel Core i7 12700KF}, 64GB RAM and used the provided graphs from the PACE challenge.
\subsection{Data Reduction} 
\begin{figure}
    \centering
    \input{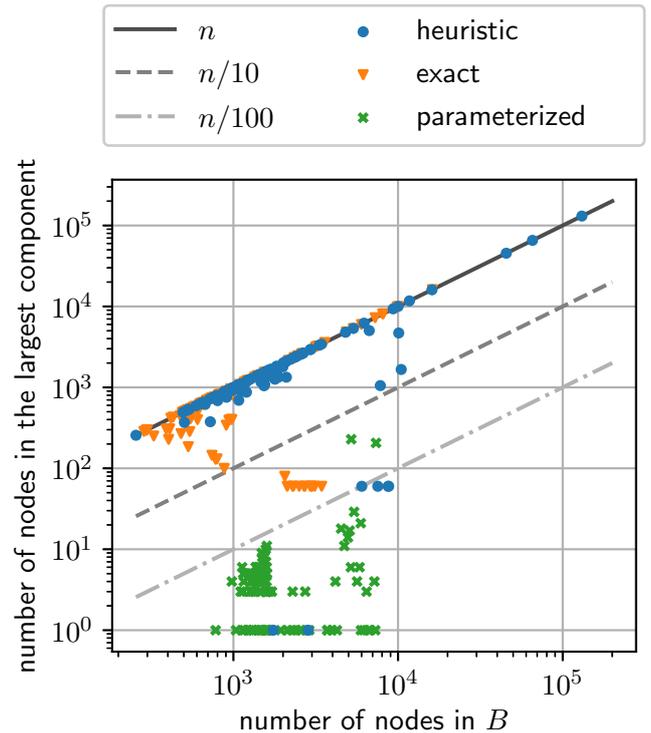}
    \caption{Showing the number of nodes in the largest component after graph splitting compared to the original number of nodes for each graph in the public dataset of PACE.}
    \label{fig:splitting}
\end{figure}

First, we want to focus on the effect of the data reduction rules we introduced in \cref{sec:data_reduction}. For this, \cref{fig:splitting} shows the number of nodes of the largest component after splitting in comparison to the number of nodes in the original graph. 

Here, we can see that on most instances of the exact and heuristic track no splitting can be applied or that at least one large component still remains. However we can also see that there are $18$ graphs in the heuristic and exact track where the largest component has less than $10\%$ of the original size or sometimes even $1\%$. 

When we look at the parameterized instances we can observe that graph splitting is really effective. For $92\%$ of the instances the resulting graph has fewer then $10$ nodes and there are even only $2$ graphs with more than $100$ nodes. This observation makes the parameterized instances with our exact solver easily solvable.

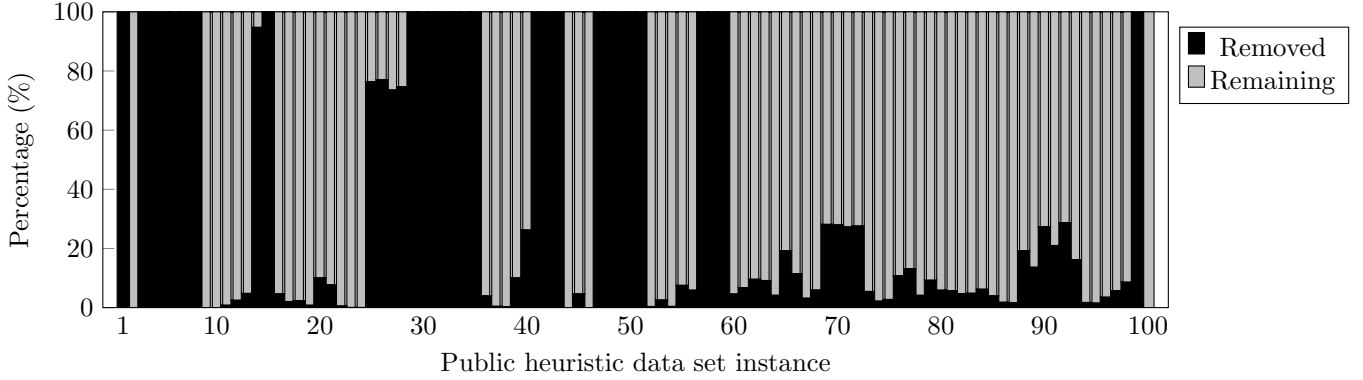
\begin{figure*}[!t]
    \centering
    \begin{tikzpicture}
    \begin{axis}[
        width=0.85\textwidth,
        height=0.3\textwidth,
        ybar stacked,
        bar width=1.2,
        xlabel={Public heuristic data set instance},
        ylabel={Percentage (\%)},
        ymin=0, ymax=100,
        xtick={1,10,20,...,100},
        xtick pos=left,
        ytick pos=left,
        legend style={at={(1.09,0.95)},anchor=north,legend columns=1},
        enlarge x limits=0.02
    ]
    \addplot[fill=black] table[x=Instance,y expr=\thisrow{Ratio_free}*100,col sep=comma] {csv_data/reduction_data.csv};
    \addplot[fill=lightgray] table[x=Instance,y expr=\thisrow{Complement_free}*100,col sep=comma] {csv_data/reduction_data.csv};
    \legend{Removed, Remaining}
    \end{axis}
\end{tikzpicture}
    \caption{Relative size of the free vertices set $B$ after graph splitting and data reduction of instances in the public heuristic data set provided by PACE.}
    \label{fig:reduction}
\end{figure*}

After we applied graph splitting, we apply the other reduction rules. To show the effect of those, we can see in \cref{fig:reduction} the number of deleted nodes after applying all reduction rules for the heuristic data. Here we can see that $27$ instances are solved using data reduction rules only and that there are $94$ instances where at least one node is removed, leaving only $6$ instances where no reduction is applied, of which $3$ are large graphs where we did not try to apply any reduction rule to, because our implementation relies on initializing a crossing matrix, which was infeasible for these large instances due to their size.

\subsection{Heuristic Solver} 

\begin{table*}[t]
    \centering
    \begin{tabular}{@{}lccccrr@{}}
            \toprule
            Algorithm & Local search & Force Swapping & SMAC & RR & Points & Time [s] \\\midrule
            \textbf{Simple approaches:}  \\
            average & & & & & $173.27989$ & $6$ \\
            median & & & & & $187.85465$ & $6$ \\
            min from median and average  & & & & & $194.18893$ & $9$ \\
            \midrule
            \textbf{Local Search models:}  \\
            no rr & x & x & x & & $199.999\textbf{29}$ & $37710$ \\
            no force swapping & x & & & x & $199.999\textbf{69}$ & $38287$ \\
            no smac & x & x & & x & $199.999\textbf{92}$ &  $38253$ \\
            submission & x & x & x & x & $199.999\textbf{98}$ & $38269$ \\
            \bottomrule
        \end{tabular}
        \caption{Points for the public and private instances according to the PACE formula.}
    \label{tab:heuristic_results}
\end{table*} 

\Cref{tab:heuristic_results} illustrates the performance of various algorithmic configurations, measured by the points awarded according to the PACE formula. For each instance, points are calculated as the ratio of crossings in the current solution to the best known solution, with the final score being the sum across all instances.

The results demonstrate that simple heuristics, such as average and median, perform significantly worse than local search approaches. While the median heuristic outperforms the average, an algorithm that selects the minimum value between median and average shows a significant improvement. This suggests that neither average nor median consistently excels across all instances.

To evaluate the impact of different features on our local search algorithm, we deactivated certain components and looked at the resulting effect on the performance of the resulting algorithm:

\begin{enumerate}
    \item Deactivating all reduction rules proved most impactful, resulting in a performance decrease of $69 \times 10^{-5}$ points compared to the final solution. 
    
    \item Removing force swapping while retaining data reduction rules led to a decrease of $29 \times 10^{-5}$ points. In this configuration, the algorithm applied data reduction rules, then iteratively selected a random solution and applied local search until the time limit was reached.
    
    \item Reactivating force swapping with our initial guessed parameters (force swap radius of 20 and step size of 1) improved performance by $23 \times 10^{-5}$ points relative to the configuration without force swapping.
    
    \item For the final submission, we optimized force swapping parameters using SMAC3. We selected 10 instances from the public dataset for which suboptimal solutions were not previously found by our algorithm. For this we used the public leader board from PACE where we can see the results of other algorithms. We let SMAC3 ran for 24 hours to optimize parameters for these instances. Implementation of these optimized parameters yielded an additional performance gain of $6 \times 10^{-5}$ points in our final solution.
\end{enumerate}

\subsection{Exact Solver}
\begin{figure}
\input{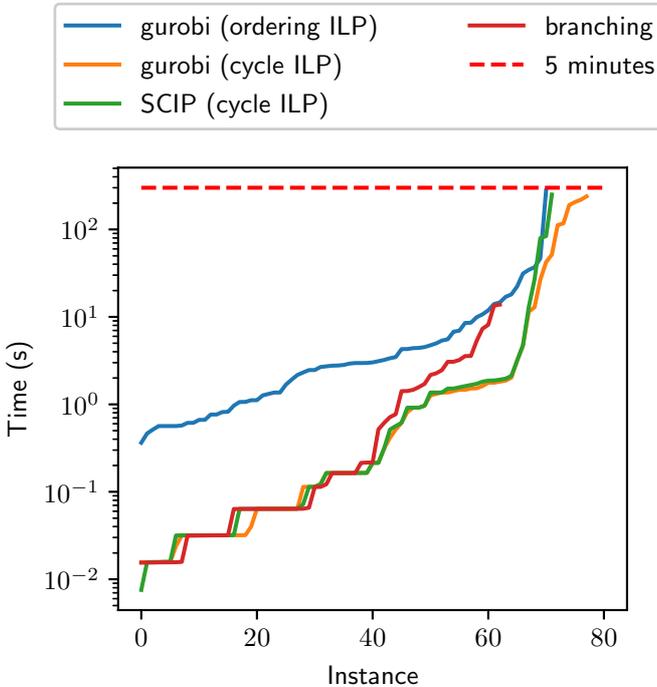}
\caption{Cactus plot comparing solution times (excluding time to calculate the heuristic) for the public exact instances.}
\label{fig:compare_exact_solver}
\end{figure}

In \cref{fig:compare_exact_solver}, we show a comparison of different exact solvers. First we compare our two ILP formulations, both of which employ row generation to improve efficiency and use Gurobi as the ILP solver. We observe that the cycle ILP successfully solves $7$ additional instances and significantly reduced computational overhead when generating solutions for easy-to-solve instances compared to the linear ordering ILP. This improvement in efficiency can be attributed to the check of transitive constraints in the ordering ILP, which is not necessary for the cycle ILP.

For the PACE challenge, where commercial solvers are not allowed, we opted for the open-source SCIP ILP solver \cite{SCIP_9} in our final submission. The SCIP solver's performance was only marginally inferior, solving $6$ instances fewer compared to the Gurobi solver.

We can also observe that the branching solver performs inferior to our ILP-based approaches, solving $15$ instances fewer than our best-performing solver. However, it is noteworthy that for smaller instances, the branching solver's performance is comparable to other solvers and lags only behind on the harder instances.
\subsection{Parameterized Solver}
Our minor adjustment to the exact solver that uses the branching algorithm from \cref{sec:branching} for instances with $upper\_bound < 10$ saves about $2$ seconds for the $100$ public parameterized instances.
As we saw in \cref{fig:splitting}, the graph splitting performs extraordinarily on instances of the parameterized track.
We are unsure whether the cutwidth gives any guarantees on how splittable a graph is using our methods in general but the parameterized instances' penalty graphs always consisted of small strongly connected components.
All of the instances had low cutwidth (the maximum was $81$) in relation to the overall instance size (in the case of the instance with cutwidth $81$ the set $A$ had $4824$ and $B$ $3452$ vertices).
Small cutwidth may be a bound for the size of strongly connected components in the penalty graph as the example in \cref{fig:ocm_cycle} already requires a cutwidth of $3$ for a simple cycle of $3$ vertices.
\subsection{PACE Results}
\begin{figure}[t]
    \centering
    \input{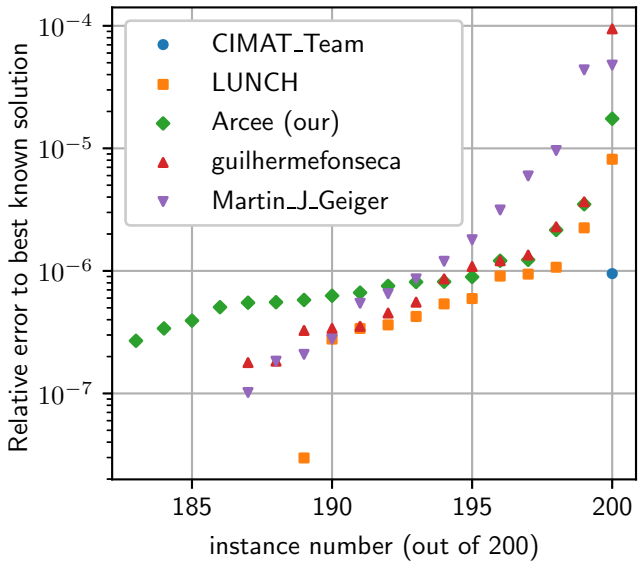}
    \caption{Cactus plot comparing the relative error to the best-known solution for the top 5 heuristics from PACE, evaluated on 200 heuristic graphs. The graphs are ordered after the average error of the solvers.}
    \label{fig:pace_heuristic}
\end{figure}

For the heuristic track, we re-evaluated the performance of the top five solvers on all 200 instances using our own hardware. The results are presented in \cref{fig:pace_heuristic}. Notably, $175$ instances yielded identical solutions across all the top solvers. In comparison to other solvers, our approach demonstrates a higher rate of non-optimal solutions. However, the average error relative to the best-known solution remains competitive leading to a good performing algorithm.

In the official rankings, our solver achieved fourth place with a score of $199.9998$ out of a maximum of $200$.

Our exact solver successfully solved $152$ out of $200$ instances, getting the eighth place in the overall solver ranking. The top-performing solver in this track, solved an impressive $199$ instances within the $30$ minute time limit.

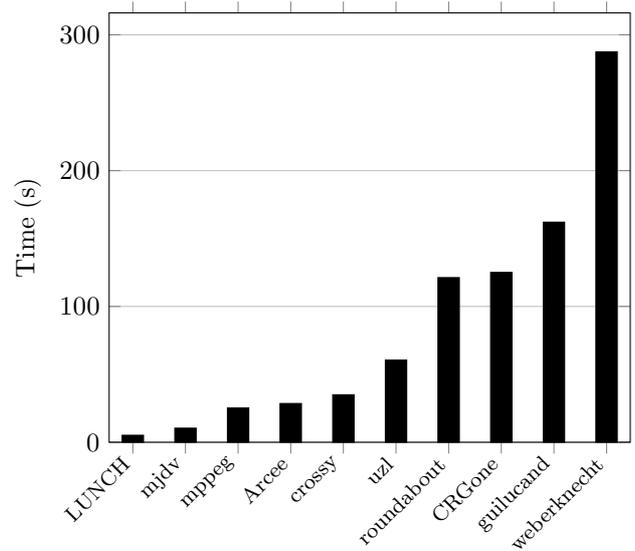
\begin{figure}[t]
        \centering
        \begin{tikzpicture}
            \begin{axis}[
                ybar,
                symbolic x coords={LUNCH, mjdv, mppeg, Arcee, crossy, uzl, roundabout, CRGone, guilucand, weberknecht},
                xtick=data,
                x tick label style={rotate=45, anchor=east, font=\footnotesize},
                ylabel={Time (s)},
                xlabel={},
                ymin=0,
                enlarge x limits=0.05,
                bar width=8pt,
                ymajorgrids,
                ]
            \addplot[
                ybar,
                fill=black
            ] table[
                x=Team Name,
                y=Time (s),
                col sep=comma,
            ] {csv_data/fpt_results.csv};
            \end{axis}
        \end{tikzpicture}
        \caption{PACE results of the parameterized track.}
        \label{fig:parameterized_times}
\end{figure}

In the parameterized track the top ten teams were able to solve all 200 instances.
Thus, running time became the deciding factor.
Our solver achieved fourth place with a total running time of $28.54$ seconds.
The ranking is visualized in \cref{fig:parameterized_times}.

The complete ranking can be found on the \href{https://pacechallenge.org/2024/results/}{official PACE website}

\section{Conclusion}
In this paper, we presented our OCM solver, Arcee, developed for the PACE 2024 Challenge. Our approach uses a combination of graph splitting techniques, data reduction rules, and both heuristic and exact solving methods to tackle the \textsc{One-Sided Crossing Minimization} problem effectively.

Our experimental results demonstrated the effectiveness of our data reduction techniques, particularly in the parameterized track where graph splitting was highly successful. The heuristic solver showed significant improvements over simple approaches like median or average heuristics, while our exact solver performed competitively, solving 152 out of 200 instances in the challenge.

In the PACE 2024 Challenge, our solver achieved the following results:
\begin{itemize}
    \item 4th place in the heuristic track (1st in the student track) with a score of $199.9998$ out of $200$
    \item 8th place in the exact track (2nd in the student track), solving $152$ out of $200$ instances
    \item 4th place in the parameterized track (1st in the student track), solving all instances with a total runtime of $28.54$ seconds
\end{itemize}

Unfortunately, we chose overly cautious limits for our categorization of small and large graphs and the running time we allowed our heuristic program. There was always an additional minute to safely shut down the solver after running over the time limit. Using this extra time in the heuristic track could have made a significant difference in the final ranking, as the scores of the best-performing heuristics were so tightly together.

In conclusion, we are proud to have presented a considerable student submission to the 2024 PACE Challenge that is on par with the best submissions to the heuristics track and parameterized track. 
\bibliography{description}

\begin{thebibliography}{27}
\providecommand{\natexlab}[1]{#1}
\providecommand{\url}[1]{\texttt{#1}}
\expandafter\ifx\csname urlstyle\endcsname\relax
  \providecommand{\doi}[1]{doi: #1}\else
  \providecommand{\doi}{doi: \begingroup \urlstyle{rm}\Url}\fi

\bibitem[Baharev et~al.(2021)Baharev, Schichl, Neumaier, and
  Achterberg]{ILPSolver}
Ali Baharev, Hermann Schichl, Arnold Neumaier, and Tobias Achterberg.
\newblock An exact method for the minimum feedback arc set problem.
\newblock \emph{{ACM} J. Exp. Algorithmics}, 26:\penalty0 1.4:1--1.4:28, 2021.
\newblock \doi{10.1145/3446429}.
\newblock URL \url{https://doi.org/10.1145/3446429}.

\bibitem[Battista et~al.(1998)Battista, Eades, Tamassia, and
  Tollis]{battista1998graph}
Giuseppe~Di Battista, Peter Eades, Roberto Tamassia, and Ioannis~G Tollis.
\newblock \emph{Graph drawing: algorithms for the visualization of graphs}.
\newblock Prentice Hall PTR, 1998.

\bibitem[Bolusani et~al.(2024)Bolusani, Besan{\c{c}}on, Bestuzheva, Chmiela,
  Dion{\'{i}}sio, Donkiewicz, van Doornmalen, Eifler, Ghannam, Gleixner,
  Graczyk, Halbig, Hedtke, Hoen, Hojny, van~der Hulst, Kamp, Koch, Kofler,
  Lentz, Manns, Mexi, M\"{u}hmer, Pfetsch, Schl{\"o}sser, Serrano, Shinano,
  Turner, Vigerske, Weninger, and Xu]{SCIP_9}
Suresh Bolusani, Mathieu Besan{\c{c}}on, Ksenia Bestuzheva, Antonia Chmiela,
  Jo{\~{a}}o Dion{\'{i}}sio, Tim Donkiewicz, Jasper van Doornmalen, Leon
  Eifler, Mohammed Ghannam, Ambros Gleixner, Christoph Graczyk, Katrin Halbig,
  Ivo Hedtke, Alexander Hoen, Christopher Hojny, Rolf van~der Hulst, Dominik
  Kamp, Thorsten Koch, Kevin Kofler, Jurgen Lentz, Julian Manns, Gioni Mexi,
  Erik M\"{u}hmer, Marc~E. Pfetsch, Franziska Schl{\"o}sser, Felipe Serrano,
  Yuji Shinano, Mark Turner, Stefan Vigerske, Dieter Weninger, and Lixing Xu.
\newblock {The SCIP Optimization Suite 9.0}.
\newblock Technical report, Optimization Online, February 2024.
\newblock URL
  \url{https://optimization-online.org/2024/02/the-scip-optimization-suite-9-0/}.

\bibitem[Chung(1985)]{cutwidth}
Fan~RK Chung.
\newblock On the cutwidth and the topological bandwidth of a tree.
\newblock \emph{SIAM Journal on Algebraic Discrete Methods}, 6\penalty0
  (2):\penalty0 268--277, 1985.

\bibitem[Djidjev and Vrt'o(2003)]{cutwidth_lower_bound_cr}
Hristo Djidjev and Imrich Vrt'o.
\newblock Crossing numbers and cutwidths.
\newblock \emph{Journal of Graph Algorithms and Applications}, 7\penalty0
  (3):\penalty0 245--251, 2003.

\bibitem[Dobler(2023)]{dobler2023note}
Alexander Dobler.
\newblock A note on the complexity of one-sided crossing minimization of trees.
\newblock \emph{arXiv preprint arXiv:2306.15339}, 2023.

\bibitem[Dujmovic et~al.(2006)Dujmovic, Fellows, Hallett, Kitching, Liotta,
  McCartin, Nishimura, Ragde, Rosamond, Suderman, et~al.]{dujmovic2006fixed}
Vida Dujmovic, Michael Fellows, Michael Hallett, Matthew Kitching, Giuseppe
  Liotta, Catherine McCartin, Naomi Nishimura, Prabhakar Ragde, Fran Rosamond,
  Matthew Suderman, et~al.
\newblock A fixed-parameter approach to 2-layer planarization.
\newblock \emph{Algorithmica}, 45:\penalty0 159--182, 2006.

\bibitem[Dujmovi{\'c} et~al.(2008)Dujmovi{\'c}, Fellows, Kitching, Liotta,
  McCartin, Nishimura, Ragde, Rosamond, Whitesides, and
  Wood]{dujmovic2008parameterized}
Vida Dujmovi{\'c}, Michael~R Fellows, Matthew Kitching, Giuseppe Liotta,
  Catherine McCartin, Naomi Nishimura, Prabhakar Ragde, Frances Rosamond, Sue
  Whitesides, and David~R Wood.
\newblock On the parameterized complexity of layered graph drawing.
\newblock \emph{Algorithmica}, 52:\penalty0 267--292, 2008.

\bibitem[Dujmovic et~al.(2008)Dujmovic, Fernau, and Kaufmann]{FPT_OCM}
Vida Dujmovic, Henning Fernau, and Michael Kaufmann.
\newblock Fixed parameter algorithms for one-sided crossing minimization
  revisited.
\newblock \emph{J. Discrete Algorithms}, 6\penalty0 (2):\penalty0 313--323,
  2008.
\newblock \doi{10.1016/J.JDA.2006.12.008}.

\bibitem[Eades and Wormald(1994)]{eades1994edge}
Peter Eades and Nicholas~C Wormald.
\newblock Edge crossings in drawings of bipartite graphs.
\newblock \emph{Algorithmica}, 11:\penalty0 379--403, 1994.

\bibitem[Gr{\"{o}}tschel et~al.(1984)Gr{\"{o}}tschel, J{\"{u}}nger, and
  Reinelt]{ILPRef1}
Martin Gr{\"{o}}tschel, Michael J{\"{u}}nger, and Gerhard Reinelt.
\newblock A cutting plane algorithm for the linear ordering problem.
\newblock \emph{Oper. Res.}, 32\penalty0 (6):\penalty0 1195--1220, 1984.
\newblock \doi{10.1287/OPRE.32.6.1195}.
\newblock URL \url{https://doi.org/10.1287/opre.32.6.1195}.

\bibitem[Gusfield(2019)]{ILP1}
Dan Gusfield.
\newblock \emph{Integer Linear Programming in Computational Biology: Overview
  of ILP, and New Results for Traveling Salesman Problems in Biology}, pages
  373--404.
\newblock Springer International Publishing, Cham, 2019.
\newblock ISBN 978-3-030-10837-3.
\newblock \doi{10.1007/978-3-030-10837-3_15}.
\newblock URL \url{https://doi.org/10.1007/978-3-030-10837-3_15}.

\bibitem[J{\"u}nger and Mutzel(2002)]{junger20022}
Michael J{\"u}nger and Petra Mutzel.
\newblock 2-layer straightline crossing minimization: Performance of exact and
  heuristic algorithms.
\newblock In \emph{Graph algorithms and applications i}, pages 3--27. World
  Scientific, 2002.

\bibitem[Kobayashi and Tamaki(2015)]{interval_graph_FPT_OCM}
Yasuaki Kobayashi and Hisao Tamaki.
\newblock A fast and simple subexponential fixed parameter algorithm for
  one-sided crossing minimization.
\newblock \emph{Algorithmica}, 72\penalty0 (3):\penalty0 778--790, 2015.
\newblock \doi{10.1007/S00453-014-9872-X}.

\bibitem[Kratsch(2014)]{ILP2}
Stefan Kratsch.
\newblock Recent developments in kernelization: {A} survey.
\newblock \emph{Bull. {EATCS}}, 113, 2014.
\newblock URL \url{http://eatcs.org/beatcs/index.php/beatcs/article/view/285}.

\bibitem[Lan et~al.(2007)Lan, DePuy, and Whitehouse]{SetCover}
Guanghui Lan, Gail~W. DePuy, and Gary~E. Whitehouse.
\newblock An effective and simple heuristic for the set covering problem.
\newblock \emph{Eur. J. Oper. Res.}, 176\penalty0 (3):\penalty0 1387--1403,
  2007.
\newblock \doi{10.1016/J.EJOR.2005.09.028}.

\bibitem[Lindauer et~al.(2022)Lindauer, Eggensperger, Feurer, Biedenkapp, Deng,
  Benjamins, Ruhkopf, Sass, and Hutter]{SMAC}
Marius Lindauer, Katharina Eggensperger, Matthias Feurer, André Biedenkapp,
  Difan Deng, Carolin Benjamins, Tim Ruhkopf, René Sass, and Frank Hutter.
\newblock Smac3: A versatile bayesian optimization package for hyperparameter
  optimization.
\newblock \emph{Journal of Machine Learning Research}, 23\penalty0
  (54):\penalty0 1--9, 2022.
\newblock URL \url{http://jmlr.org/papers/v23/21-0888.html}.

\bibitem[Matuszewski et~al.(1999)Matuszewski, Sch{\"{o}}nfeld, and
  Molitor]{OCM_sifting}
Christian Matuszewski, Robby Sch{\"{o}}nfeld, and Paul Molitor.
\newblock Using sifting for k -layer straightline crossing minimization.
\newblock In Jan Kratochv{\'{\i}}l, editor, \emph{Graph Drawing, 7th
  International Symposium, GD'99, Stir{\'{\i}}n Castle, Czech Republic,
  September 1999, Proceedings}, volume 1731 of \emph{Lecture Notes in Computer
  Science}, pages 217--224. Springer, 1999.
\newblock \doi{10.1007/3-540-46648-7\_22}.

\bibitem[Nagamochi(2005{\natexlab{a}})]{OCM_lower_bound}
Hiroshi Nagamochi.
\newblock On the one-sided crossing minimization in a bipartite graph with
  large degrees.
\newblock \emph{Theor. Comput. Sci.}, 332\penalty0 (1-3):\penalty0 417--446,
  2005{\natexlab{a}}.
\newblock \doi{10.1016/J.TCS.2004.10.042}.

\bibitem[Nagamochi(2005{\natexlab{b}})]{nagamochi2005improved}
Hiroshi Nagamochi.
\newblock An improved bound on the one-sided minimum crossing number in
  two-layered drawings.
\newblock \emph{Discrete \& Computational Geometry}, 33:\penalty0 569--591,
  2005{\natexlab{b}}.

\bibitem[Pho and Lapidus(1973)]{ILPRef2}
TK~Pho and L~Lapidus.
\newblock Topics in computer-aided design: Part i. an optimum tearing algorithm
  for recycle systems.
\newblock \emph{AIChE Journal}, 19\penalty0 (6):\penalty0 1170--1181, 1973.

\bibitem[Rudell(1993)]{rudell1993dynamic}
R.~Rudell.
\newblock Dynamic variable ordering for ordered binary decision diagrams.
\newblock In \emph{Proceedings of 1993 International Conference on Computer
  Aided Design (ICCAD)}, pages 42--47, 1993.
\newblock \doi{10.1109/ICCAD.1993.580029}.

\bibitem[Sechen(2012)]{sechen2012vlsi}
Carl Sechen.
\newblock \emph{VLSI placement and global routing using simulated annealing},
  volume~54.
\newblock Springer Science \& Business Media, 2012.

\bibitem[Stallmann et~al.(2001)Stallmann, Brglez, and
  Ghosh]{stallmann2001heuristics}
Matthias Stallmann, Franc Brglez, and Debabrata Ghosh.
\newblock Heuristics, experimental subjects, and treatment evaluation in
  bigraph crossing minimization.
\newblock \emph{Journal of Experimental Algorithmics (JEA)}, 6:\penalty0 8--es,
  2001.

\bibitem[Sugiyama et~al.(1981)Sugiyama, Tagawa, and Toda]{PenaltyGraph}
Kozo Sugiyama, Shojiro Tagawa, and Mitsuhiko Toda.
\newblock Methods for visual understanding of hierarchical system structures.
\newblock \emph{{IEEE} Trans. Syst. Man Cybern.}, 11\penalty0 (2):\penalty0
  109--125, 1981.
\newblock \doi{10.1109/TSMC.1981.4308636}.

\bibitem[Toth(2000)]{ILP3}
Paolo Toth.
\newblock Optimization engineering techniques for the exact solution of np-hard
  combinatorial optimization problems.
\newblock \emph{European Journal of Operational Research}, 125\penalty0
  (2):\penalty0 222--238, 2000.
\newblock ISSN 0377-2217.
\newblock \doi{https://doi.org/10.1016/S0377-2217(99)00453-1}.
\newblock URL
  \url{https://www.sciencedirect.com/science/article/pii/S0377221799004531}.

\bibitem[Waddle and Malhotra(1999)]{Segtrees}
Vance~E. Waddle and Ashok Malhotra.
\newblock An {E} log {E} line crossing algorithm for levelled graphs.
\newblock In Jan Kratochv{\'{\i}}l, editor, \emph{Graph Drawing, 7th
  International Symposium, GD'99, Stir{\'{\i}}n Castle, Czech Republic,
  September 1999, Proceedings}, volume 1731 of \emph{Lecture Notes in Computer
  Science}, pages 59--71. Springer, 1999.
\newblock \doi{10.1007/3-540-46648-7\_6}.
\newblock URL \url{https://doi.org/10.1007/3-540-46648-7\_6}.

\end{thebibliography}
\end{document}